\documentclass[12pt,preprint]{aastex}
 \usepackage{natbib}
 \usepackage{epsfig}
\revised{--}
\accepted{--}
\cpright{}{2003}
\slugcomment{Accepted for publication in Ap. J.}
\shorttitle{Detection of a white light CME shock}
\shortauthors{Vourlidas et al.}
\received{2003 March 7}
\begin{document}
\title{Direct Detection of a CME-Associated Shock in LASCO White Light
Images} 
\author{A. Vourlidas}
\affil{Naval Research Laboratory, Washington, DC 20375} 
\email{vourlidas@nrl.navy.mil}

\author{S.T. Wu and A.H. Wang}
\affil{CSPAR \& Dept. of Mech. \& Aero. Eng.,
University of Alabama in Huntsville, Huntsville, AL 35899}

\author{P. Subramanian}
\affil{IUCAA, P.O. Bag 4, Ganeshkind, Pune 411007, India}
\and

\author{R.A. Howard}
\affil{Naval Research Laboratory, Washington, DC 20375} 
\begin{abstract}
The LASCO C2 and C3 coronagraphs recorded a unique coronal mass
ejection on April 2, 1999. The event did not have the typical
three-part CME structure and involved a small filament eruption
without any visibile overlying streamer ejecta. The event exhibited an
unusually clear signature of a wave propagating at the CME flanks. The
speed and density of the CME front and flanks were consistent with the
existence of a shock. To better establish the nature of the white
light wave signature, we employed a simple MHD simulation using the
LASCO measurements as constraints.  Both the measurements and the
simulation strongly suggest that the white light feature is the
density enhancement from a fast-mode MHD shock. In addition, the LASCO
images clearly show streamers being deflected when the shock impinges
on them. It is the first direct imaging of this interaction.
\end{abstract}
%
\section{Introduction}

White light coronagraphs regularly observe fast, impulsive coronal
mass ejections (CMEs) with speeds that exceed both the coronal sound
speed ($\sim200$ km/s) and the Alfv\'{e}n speed \citep[$\sim800$ km/s
at 4 R$_{\sun}$][]{mann99}. Therefore, CMEs are capable of
driving wave perturbations and even shocks in the low corona
\citep[e.g.,][]{hund88}. However, the imaging of CME-associated
shocks remains an observational challenge. There are two observational
approaches to this problem.

The first approach relies on the study of shock proxies. The existence
of shocks during mass ejections is supported by a large amount of
indirect evidence, such as radio type-II observations
\citep[e.g.,][]{cliver99} and distant streamer deflections
\citep{gosling74,michels84, sheeley00}. The contribution of type-II
observations to the the study of coronal CME shocks is limited by the
lack of imaging of type-II sources, the frequent concurrence of CMEs
and flares (flares can also drive shocks in the low corona), and the
uncertainty in the CME initiation times. Although there are some
promising recent results \citep{maia00}, the connection between
type-IIs and CMEs will likely remain unclear for the near
future. Observations of distant streamer deflections offer the best,
indirect, evidence for white light shocks. Such observations have been
more numerous in recent years thanks to the increased sensitivity and
temporal coverage of the LASCO coronagraphs \citep{sheeley00}.

The second approach to the problem is to establish shock signatures
directly from the white light CME images. Initially, it was thought that
the looplike front of many CMEs was the density enchancement from a
fast MHD shock \citep[e.g.,][]{maxwell81,steinolfson85}. But
\cite{sime84} pointed out several discrepancies between the model
predictions and the observations. They argued, for example, that many
looplike CMEs propagate too slowly to form a fast shock. This led
\cite{hund87} to suggest slow shocks as candidates for some CME
fronts. \cite{steinolfson90a,steinolfson90b} simulated in some detail
the appearance of both slow and intermediate shocks on coronagraph
images. However, the applicability of the simulations to the CME
analysis is hindered by the complexity of the ejected structures. Many
CMEs have irregular fronts or no well-defined fronts at all and it is
usually difficult to differentiate between coronal material, which is
inherently looplike, and shock-related structures.

In fact, there has been only one published identification of a white
light shock despite the observations of thousands of CMEs since the
early 1970s. The observation of an unusual looplike CME with the {\it
Solar Maximum Mission\/} (SMM) coronagraph \citep{mcqueen80} led
\cite{sime87} to suggest that the loop front could be a fast shock
wave. They argued that the high lateral expansion speed (800 km/s) of
the CME and the absence of any deflections of coronal structures before
the loop impinged on them were strong indicators of a shock. There
were some problems with this interpretation, however, that arose from
the limitations of the available observations. For example, the
restricted field of view and sensitivity of the SMM images could not
rule out the existence of a disturbance ahead of the observed CME nor
could provide a reliable determination of the acceleration profile of
the CME. One could only note that the loop did not appear to
deccelerate during the streamer crossings contrary to theoretical
\citep{odstr00} and observational \citep{mcqueen83,sheeley00}
expectations. Finally, the nature of the loop front could not be
established without observations in other wavelengths.

The high quality LASCO observations provide an excellent dataset for
locating possible white light signatures. We searched the LASCO
database for CME events which were simple enough to allow an
unambiguous identification of a white light shock. The best case is an
event on April 2, 1999, when a fast ejection with an exceptionally
clear density enhancement along its flanks was observed. The
availability of CME simulation codes and high quality datasets, in
several wavelengths, provide us with the necessary tools to
investigate the nature of the observed enchancement more thoroughly
than it was possible in the past. We start with an overview of the
available observations in \S~2. The results from the analysis of the
observations and the MHD modeling are presented in \S~3 and we discuss
their implications in \S~4. We summarize our findings in \S~5.

\section{Observations}

On April 2, 1999, the Large and Spectroscopic Coronagraph
\citep[LASCO,][]{brueckner95} and the Extreme Ultraviolet Imaging
Telescope \citep[EIT,][]{boudine95} aboard the Solar and Heliospheric
Observatory \citep[SOHO,][]{domingo95} observed a jet-like CME
(jet-CME hereafter) at the northeast solar limb (Figure 1). Ejections
of this type are characterized by their small widths
($\sim10^\circ-20^\circ$), limited latitudinal expansion and simple
structure \citep{gilbert01,dobrzycka03}. We can easily identify the source region
in the EIT images because part of the ejecta can be seen in absorption
in the 195\AA\ image (Figure 1). The jet-CME originated in NOAA active
region 8507.  An M1.1 soft X-ray flare was reported from the same
location (Solar Geophysical Data Reports). The flare began at 8:03 UT,
and peaked at 8:21 UT.  The EIT images between 8:12 -- 8:54 UT showed
a series of dark spray ejections (Figures 1-2). By 8:30 UT, the front
of the jet-CME appeared in the C2 field of view at a height of 3.5
R$_{\sun}$\ (Figure 2). The ejection maintained its narrow width
($\sim20^\circ$) out to 30 R$_{\sun}$. The filamentary structure of
the ejecta clearly points to a filament eruption in association with
the jet-CME. A filament was visible in the H$_\alpha$ image from the
Meudon observatory taken at 8:17 UT. This is not a typical filament
eruption because the event lacks the familiar loop-cavity-core
configuration. This is apparent upon inspection of the C2 image, which
illustrates a loop-cavity-core shape CME occurring over the western
limb (Fig.~1). The western CME has a well-defined loop front followed
by a cavity and what appears to be a core just over the C2
occulter. By comparison, the jet-CME event appears to consist of the
main body of the CME without a loop or cavity preceeding the ejecta.
We can think of two possible explanations for the lack of overlying
streamer material in the jet-CME:
\begin{itemize}
\item The jet-CME occurred outside a streamer which is not an
uncommon situation. In a study of a large sample of LASCO CMEs,
\cite{subramanian99} found that 27\% of the
events were displaced from preexisting streamers.

\item The overlying corona was disrupted by an earlier CME (hereafter
CME1) that was first seen in C2 at 1:30 UT and had just exited the C2
field of view when the jet-CME followed in its wake. The apparent
position angle of CME1 was $91^\circ$, close to the position angle of
the jet-CME ($72^\circ$). CME1 was also wide enough ($74^\circ$) to
extend over the location of the jet-CME. We carefully inspected the
LASCO-C2 images to look for signatures of coronal depletions after
CME1. The only significant brightness depressions were localized over
the location of CME1. However, this result does not preclude the
possibility that most of the material overlying the jet-CME was
located below $\sim 2.3$ R$_{\sun}$, the inner cutoff of the C2
coronagraph. In that case, no depletions would be visible in the LASCO
images. Such a coronal configuration is also consistent with the lack
of any obvious relation of AR8507 to a C2 streamer during its east
limb passage. Therefore, the observations do not dismiss the
possibility that CME1 carried away part of the streamer material over
the site of the jet-CME, leading to the unusual appearance of the
jet-CME.
\end{itemize}

The most intriguing aspect of this event, however, is the sharpness of
the southern flank of the CME. A much fainter counterpart is barely
visible along the northern flank. The CME bears an uncanny resemblance
to a fast projectile and its associated bow shock. This similarity
prompted our investigation on whether the sharp CME flank is actually
the white light counterpart of a shock.

We searched the available datasets in other wavelengths for possible
shock evidence in association with this CME. No metric type-II
emission, the main proxy for coronal shocks, was reported by any radio
spectrographs. Our own analysis of the Potsdam spectrograms revealed
only a group of type-III emissions between 8:09-8:15 UT, which was
most likely connected to the flaring in the active region.  The
Ultraviolet Coronagraph Spectrometer \citep[UVCS,][]{kohl95} was
observing along the northeast limb between 8:11 and 9:44
UT. Unfortunately, the UVCS slit did not intercept the CME flanks but
it intercepted part of the filament in the CME core. The spectra
suggest that the core is an untwisting filament (A. Chiaravella,
personal communication). The \textit{Transition Region and Coronal
Explorer\/} \citep[TRACE,][]{handy99} 171\AA\ and the
\textit{Yohkoh\/}/Soft Xray Telescope \citep{tsuneta91} images, during
this event, do not show any coronal features or ejecta that could be
identified with the sharp white light flank. 

We conclude that these observations do not provide strong evidence
either for a shock or for a coronal structure that could be
associated with the white light feature. To establish the nature of
the white light feature, therefore, we rely on the analysis of the
LASCO data in conjuction with an MHD simulation.

\section{Analysis of the White Light CME}

\subsection{LASCO Measurements}

We analysed the kinematics of the event by constructing height-time
plots along two radial positions using the EIT, C2 and C3 images. The
first position angle (PA), marked as P1 in Figure 2, corresponds to
the front of the filament and consequently of the CME
(PA=80$^\circ$). The second position (P2) was taken at a random
position along the southern CME flank (PA=100$^\circ$). The position
angles were measured counterclockwise from the solar North
Pole. Figure 3 demonstrates that the height-time curves can be fitted
well by second degree polynomials. Both curves show deceleration. The
average speeds in both locations (800-1000 km/s) are well above the
median speed of 476 km/s for CMEs observed with LASCO
\citep{stcyr00}. They are comparable to shock speeds deduced from
metric type-II bursts \citep[e.g.,][]{klassen00} and are similar to
the measured speeds for the July 6, 1980 white light shock candidate
\citep{sime87}. The derived speeds, therefore, are consistent with the
existence of a shock. Note that, as all coronagraphic measurements,
the above heights, speeds and decelerations are projected values in
the plane of the sky. In this case, however, we are confident that the
measured parameters are close to their true, radial values because the
CME source region is very close to the solar limb and since most CMEs
appear to move radially \citep{hund94,stcyr99,stcyr00} the jet-CME was
most likely propagating close to the plane of the sky.

We measured the mass across the sharp feature and derived the
corresponding density profile, as follows: First, the LASCO C2 and C3
images were corrected for vignetting, exposure time and other
instrumental effects and were calibrated in units of excess brightness
after subtracting a pre-event image. This is a standard procedure for
the analysis of coronagraph CME observations \citep{poland81} and, for
the LASCO case, is described in some detail in \cite{vour00}.  The
resulting excess brightness images were converted to excess mass
images under the usual assumptions: (i) the scattering electrons are
concentrated on the plane of the sky and, (ii) the ejected material
comprises a mix of completely ionized hydrogen and 10\% helium. These
 images are shown in Figure 2. The visibility of the CME flank is
enhanced somewhat by this procedure and it can be followed out to at
least 20 R$_{\sun}$. We measured the mass profiles across the front at
the same position angle (P2) where the height-time measurements were
taken. To improve the signal-to-noise ratio, especially at the larger
elongations, we averaged over a 10$^\circ$-wide swath along the
flank. The result is a measurement of the line of sight density of the
white light structure. To convert the profiles to volume density we
made another assumption about the unknown depth of the feature along
the line of sight. The sharpness of the feature suggests a limited
extent along the line of sight. Therefore, we assumed its depth equal
to its width ($\sim0.2$ R$_{\sun}$), as measured in the C2 image at
8.30 UT. The resulting profiles represent estimates of excess
density. We converted them to total density profiles by adding the
pre-event (background) coronal density. This density was derived from
the closest C2 polarization brightness (pB) image using the well-known
pB inversion method \citep{hulst50,hayes01}. The pB image was obtained
at 21:00 UT on April 1, before the passage of the earlier CME
discussed in the previous section. It is likely that the derived
background density might be slightly higher than the actual density at
the time of our event. The final density profiles are shown in
Figure~4 where they are compared to the commonly-used SPM background
model \citep{saito77} coronal density profile. We see that the
background density profile, derived from the C2 pB image, agrees very
well with the SPM model profile. The density increase across the
feature is very sharp in the C2 images and the profiles are very
similar to the expected density profile across a shock.  The profile
retains its sharpness until 9:18 UT, becoming smoother at larger
distances (not shown here). The density jump across the profile is
about a factor 3 (at 8:30 UT) and strongly suggests that the sharp
white light feature is actually a shock.  However, one should note
that, given the assumptions used to derive the density profiles, this
density jump is only an estimate. To examine the viability of a shock
at the flanks of the CME, we simulate the event using the measured
speeds and background density profiles as constraints.

\subsection{MHD Simulation}

We start with a time-dependent, 2D plane-of-sky ideal MHD model. The
model is described by the standard equations of MHD theory
\citep[e.g.,][]{priest82} with additional momentum and heating terms
to accommodate the bimodal (fast, slow) solar wind
\citep{wang98,wu00}.  The complete set of equations is given in
\cite{wang98}. The model incorporates four constraints that are based
on the LASCO observations. Namely, (i) the CME does not have a loop or
a cavity preceeding its core, (ii) the CME center is located at
80$^\circ$\ counterclockwise from the solar north pole, (iii) the event
shows significant deceleration and (iv) the simulated density profiles
should match the observed ones (Figure~4). Finally, we are not
interested in the details of the initiation of the event and thus we
use a simple initiation mechanism which is described later.

To accommodate the first constraint we need to consider the initial
magnetic field topology. Simulations with closed field configurations
and beta ratios of the order of unity tend to lead to loop-cavity-core
CME morphologies with considerable latitudinal extent while low beta,
open field configurations result in elongated, laterally confined
ejections \citep[see Figure 5 and Table 1
in][]{steinolfson78}. Therefore, closed field configurations are
better suited for modeling CMEs similar to the west limb event in
Figure~1 or to the loop-like CME described in \cite{sime87}. Instead,
the narrow width of the CME and the absence of the 3-part structure,
led us to the choice of a radial magnetic field (Figure 5) with beta
ratios ranging from 0.83 at the equator to 0.34 at the poles. Figure 5
also shows the initial solar wind velocity and density distribution at
the position angle of the CME core (P1). The model density is in good
agreement with the measured density. The background flow velocity is
obtained from the MHD model. Note that the contrast between the CME
center and polar background flow velocities is insignificant because
we use a radial magnetic field configuration. To account for the
observed deceleration, we used a spatially dependent polytropic index
($\gamma= 1.05$ between 1-4 R$_{\sun}$, then linearly increasing to
1.45 at 30 R$_{\sun}$). This implies that the energy equation for the
ideal MHD model is modified to include non-adiabatic processes
\citep{wang98}. To reduce computation time, we assumed symmetrical
conditions at the angular computational boundaries at 0$^\circ$\ and
80$^\circ$\ from solar north. In other words, the CME is considered
symmetric relative to its center at P1 which is consistent with the
appearance of the CME in the LASCO images. The lower radial boundary
is the solar surface (1 R$_{\sun}$) and is prescribed according to the
theory of characteristics \citep{wu87}. The initial conditions at the
solar surface at locations P1 and P2 are: magnetic field 1.5 G and 1.2
G, number density $5 \times 10^7$ cm$^{-3}$ and $10^8$ cm$^{-3}$, and
temperature $2.24 \times 10^6$ K and $1.74 \times 10^6$ K,
respectively.  A detailed mathematical representation of the
compatibility relations obtained by methods of characteristics in two
dimensions is given in \cite{suess96}, for example. In short, we
specify that $B_r$ and $p/\rho^\gamma$ are constants at the lower
boundary (the solar surface), and $p$, $B_\theta$, $\rm{v}_r$ and
$\rm{v}_\theta$ are computed from the compatibility relations
\citep{wu87,wang92,suess96}. The linear extrapolation method is used
to compute the upper radial computational boundary at 30 R$_{\sun}$
which has non-reflecting boundary conditions. The system is in
quasi-steady equilibrium in its initial state. The same modeling
method was applied in the simulation of another event in \cite{wu99}.

The sole objective of our simulation is the identification of the
nature of the flanks of this particular jet-CME, not the processes of
CME initiation. Thus, we do not seek elaborate initiation
configurations that attempt to model CMEs in more realistic ways but
impose a premium on computational time. Instead, we initiate the event
with a simple pressure pulse. In any case, the nature of the CME
flanks should not depend on the details of its initiation
mechanism. We introduce a 20$^\circ$-wide pressure perturbation by
raising the density but keeping the temperature identical to the local
value at the solar surface, which is symmetric relative to the center
of the event (P1). This density pulse is three times higher than the
background and the maximum perturbed velocity at its center reaches
200 km/s at 200 seconds, maintains this value for about 5000 seconds,
and then declines to the background velocity. This perturbation
represents a mass flow that forms as an ejecta to produce the observed
shock.  The total computational time is 293 minutes.

Figure 4 demonstrates the excellent agreement between the simulated
and measured density profiles. The simulated and measured height-time
curves are compared in Figure 6. Overall the simulation reproduces the
CME measurements well. We are also interested in reproducing the
morphology of the event. In Figure~7, we assemble the observed images
together with simulated pB images and the magnetic field
configuration. The core of the CME is immediately recognized acting as
the piston for the bow-shock feature.  We see that the choice of
radial field led our model to a good match of the narrow width of the
CME. 

\subsubsection{Analysis of Simulation Results}

The next step is the analysis of the waves produced in the
simulation. In an MHD medium, the information from a disturbance can
propagate with one of two characteristic speeds; the slow-mode, $V_s$
or the fast-mode, $V_f$ speed that are given by
\begin{equation}
V_{f,s}^2= 0.5\left[(V_\alpha^2 + C_s^2) \pm
\sqrt{(V_\alpha^2+C_s^2)^2-
4V_\alpha^2C_s^2\cos^2\theta}\right]  
\end{equation}
where $V_\alpha$ is the Alfv\'{e}n speed
($V_\alpha=B/\sqrt{4\pi\rho}$), $C_s$ is the sound speed ($C_s=(\gamma
p/\rho)^{1/2}$), and $\theta$ is the angle between the magnetic field
and the direction of the wave propagation. For propagation parallel to
the magnetic field ($\theta=0$), we find that $V_f=V_\alpha$ and $V_s
= C_s$. In the case of propagation perpendicular to the magnetic
field, only the fast-mode propagates with speed $V_f= (V_\alpha^2+
C_s^2)^{1/2}$.

When the wave is compressive, it may be observed in the coronagraph
images if the density increase at the wave front is large enough. For
sufficiently energetic drivers, the wave may steepen into a
shock. Note that all MHD shocks are compressive shocks. From
shock theory \citep{jeffrey64}, the existence of a MHD shock solution
for propagation parallel to the field requires
\begin{eqnarray}
 V_f>U_{sh}>V_s,& \qquad \rm{slow\ MHD\ shock}\nonumber\\
 U_{sh}>V_f,& \qquad \rm{fast\ MHD\ shock}
\end{eqnarray}
and for propagation perpendicular to the magnetic field requires
\begin{eqnarray}
U_{sh}> (V_\alpha^2+C_s^2)^{1/2} \nonumber
\end{eqnarray}
where $U_{sh}$ is the normal component of the relative shock speed
\begin{eqnarray}
U_{sh}=(\vec{V}_{prop}-\vec{V}_{SW})\cdot \vec{e} 
\end{eqnarray}
where $\vec{e}$ is the normal to the shock front, $V_{prop}$ is the
propagation speed of the disturbance and $V_{SW}$ is the solar wind
speed. In other words, $V_{prop}$ is the usual speed derived from the
coronagraph measurements and $U_{sh}$ is the speed of the disturbance
in the solar wind frame. To derive the propagation speed from the
simulation we follow \cite{wu96} and identify the locations of the
shock ($r_1,r_2,...,r_n$) at times $t_1,t_2,...,t_n$. Then
\begin{eqnarray} 
V_{prop} = {{r_{i+1}-r_i}\over{t_{i+1}-t_i}}
\end{eqnarray}
The simulation results for $U_{sh}$, $V_{f,s}$, $V_{SW}$ and
$V_{prop}$ are shown in Tables 1 and 2 for both the CME flank and
front, respectively. The characteristic speeds are calculated at a
point in the pre-shock region. We see that the wave velocity is larger
than the fast-mode speed at both locations. Therefore, our simulation
supports the existence of a shock at the flanks of the CME. But could
this shock be visible in the white light images? We can check the
visibility of the shock by estimating the density increase across the
shock front as follows. For the fast-mode shock the Mach number is
defined as
\begin{eqnarray}
M_\alpha = {{U_{sh}}\over{V_f}}
\end{eqnarray}
The maximum density increase across the shock front occurs for
propagation along the magnetic field
\begin{eqnarray}
{{\rho_2}\over{\rho_1}} = {{(\gamma+1)M_\alpha^2}\over{(\gamma-1)M_\alpha^2+2}}
\end{eqnarray}
where $\rho_1$,$\rho_2$ are the upstream and downstream densities,
respectively. For propagation at any other direction, the density
compression is smaller than Eq.(6) and there is no exact formula for
it. Only for very large Mach numbers, the density compression reduces
to
\begin{eqnarray}
{{\rho_2}\over{\rho_1}} = {{\gamma+1}\over{\gamma-1}}
\end{eqnarray} 
for any propagation direction. From Eqs (5) and (6) we can now
calculate the maximum density compression for the shock. The results
are shown in Tables 1-2 for the CME flank and front, respectively. It
may be noted that we use the polytropic index $\gamma$, and not the
ratio of specific heats, in computing the density compression
ratios. The results are close to the measured density compression,
taking into account that the actual shock does not propagate along the
magnetic field. The actual compression and Mach numbers should be less
than the predictions in Tables 1-2.

In summary, the simulation leads us to the following conclusions:
\begin{itemize}
\item The observed bow shaped feature at the flank of the CME is the
density enhancement from a fast mode MHD shock.
\item The shock strength measured by its Mach number remained high
throughout its propagation in the C2 and C3 fields of view.
\item The observed and model density enhancements (Fig.~4) show very
good agreement. The measured density compression ratios are slightly
lower than those predicted by the shock jump condition. This is
expected because the shock is propagating at an oblique angle to the largely
radial magnetic field, as an inspection of the images suggests. 

\item The deceleration of the shock and the front is simulated by
using a spatially dependent polytropic index. We do not model the
energy dissipation processes that result in the deceleration; instead,
we use a spatially varying polytropic index as a proxy for the energy
dissipation.  We have used the value of the spatially varying
polytropic index at the location of the shock in computing the density
compression ratio. In the absence of an explicit energy equation for
the flow and knowledge about radiation losses at the shock, we feel
that this is a reasonable approach. If we did solve an explicit energy
equation for the flow, we would use the ratio of specific heats in
order to compute the ratios of various quantities across the
shock. However, this too would not be correct if the shock were lossy;
one would then have to employ a separate energy balance equation
across the shock that takes these losses into account.

\end{itemize}

\section{Discussion}

As we pointed out in the introduction, it is generally difficult to
differentiate between the white light signatures of shocks (or waves)
and those of coronal ejecta such as loops and filaments.  For the
event we examined here, we can exclude the possibility that the flanks
of the CME correspond to coronal structures for several reasons.

First, there is no visible ejecta ahead of the jet-CME. The absence of
overlying material might be due to an earlier, larger CME event that
entered the C2 field of view at 1:30 UT. The earlier CME propagated
along the same path as our event and possibly carried away some of the
overlying corona. It is likely that the coronal density above the east
limb was still low when the second eruption occurred.  

Second, there is no evidence for preexisting or erupted large coronal
loops (on scales comparable to the white light front) in the
observations of every available wavelength, from EUV to X-rays.

Third, the flanks of the CME diffused rapidly as they propagated
outwards. This behaviour is inconsistent with that of a coronal
structure. This is readily apparent from a comparison to the
filamentary structures in the CME core that remain identifiable at
large heliocentric distances.  

Fourth, the southeastern streamer is clearly deflected when the CME
flank reaches its location. We then see the deflection propagating to
the adjacent streamers thus creating a "front" traveling southwards
with a slower speed than the CME front or its flanks (Figures
1-2). This behavior is expected from a shock on theoretical grounds
\citep{odstr00,holst02}. It is, also, quite common in LASCO CME images
where it has been interpreted as an indirect signature of shock or
waves caused by high-speed CMEs \citep{sheeley00}.

From the above arguments, we deduce that the sharp white light feature
at the flank of the CME is likely a wave.  Whether it is actually a
shock or not depends on the local coronal conditions. However, no
direct measurements of the magnetic field or the coronal density
exist. Thus we employed a simulation (\S3.2) to assess
the likelihood of a shock. As the high CME speeds suggest and the
results of the simulation support, the CME flank is likely the
signature of a fast mode MHD shock. The lack of metric type-II
emission does not contradict our conclusion since metric type-II
bursts do not always accompany fast CMEs and vice versa
\citep[e.g.,][]{cliver99,gopal00}.

To our knowledge, this event is the best case for a direct
identification of a white light shock in a coronagraph image. It
doubles the number of CME shock candidates to date \citep[this
work]{sime87}. Clearly, this is a very small sample and does not
permit any conclusive arguments on whether the CME fronts or other CME
features seen in LASCO images are indeed shocks. It is important to
extend this work with further studies. There are many fast LASCO CMEs
that exhibit sharp features, along either their fronts or their flanks
(Fig.~8). Based on the present work, it is likely that these features
could be shock or fast mode wave signatures.  This information might
be proven useful in the analysis of these events by providing, for
example, the possible locations of type-II sources.

Another important result, in our opinion, is the simultaneous
observation of a streamer deflection and the shock wave responsible
for it. Although, streamer deflections have always been interpreted as
the indirect signature of a CME-associated disturbance, the ``hard''
evidence was missing, until now. It is apparent, that the disturbance
propagates through the streamer at a lower speed than in the open
corona. This is most likely due to the high streamer density but
projection effects may come into play \citep{sheeley00} and would
probably complicate the derivation of any physical parameters about
the streamer (or the wave).  Although it could yield interesting
results on the properties of shocks and streamers, no such analysis
has been carried out so far, to our knowledge.

Another noteworthy consideration is the scale of the event. We showed
that apparently small spray-like ejections in the low
corona can give rise to a CME with significant influence over a much
larger area. Yet, the brightest part of the CME (its massive core) is
constrained within a narrow width ($<20^\circ$). If this was an
earth-directed event, it would have remained inside the coronagraph
occulters and hence would not have been detected. These arguments
raise the question whether there exists a class of undetected
earth-directed CMEs that can have some influence on the terrestrial
space weather conditions. This question cannot be answered until
observations of CMEs along the sun-earth line become available during
the planned STEREO mission.

\section{Conclusions}

We have analysed a unique CME which exhibited a sharp feature at its
flanks reminiscent of shock signature. From the analysis of the LASCO
data and other available observations, we concluded that the white
light flanks could not be coronal structures. We employed a MHD
simulation to assess the possibility of the flanks being the density
enhancement from a shock. We used the measured speeds and coronal
densities as constraints to the simulation and showed that a fast MHD
shock was formed at the front and the flank of the CME. We conclude
that shocks could be directly observed in white light coronagraph
images, under suitable conditions.

The event also provided us with the first direct observation of a
streamer deflection by a fast mode shock, thus lending firm support to
the interpretation of the commonly-seen streamer deflections as
proxies to CME-induced coronal shocks. Although the appearance of this
CME is rather unique, it is not fundamentally different that the
ejections seen daily in the LASCO images. Therefore, the
aforementioned results could be used in the search for shocks
signatures in other events.
\acknowledgments 

We thank the referee for helpful comments that improved the
manuscript. We thank H. Aurass for providing the Potsdam
radioheliograph data and M. Andrews for useful comments. A.V
acknowledges support from NASA grants. S.T.W and A.H.W wish to
acknowledge support by NSF grant ATM-0070385 and AFOSR grant
F49620-00-0-0204. SOHO is an international collaboration between NASA
and ESA. LASCO was constructed by a consortium of institutions: the
Naval Research Laboratory (Washington, DC, USA), the
Max-Planck-Institut fur Aeronomie (Katlenburg- Lindau, Germany), the
Laboratoire d'Astronomie Spatiale (Marseille, France) and the
University of Birmingham (Birmingham, UK).

\newpage

\clearpage

\newpage
\begin{figure}
\epsfig{file=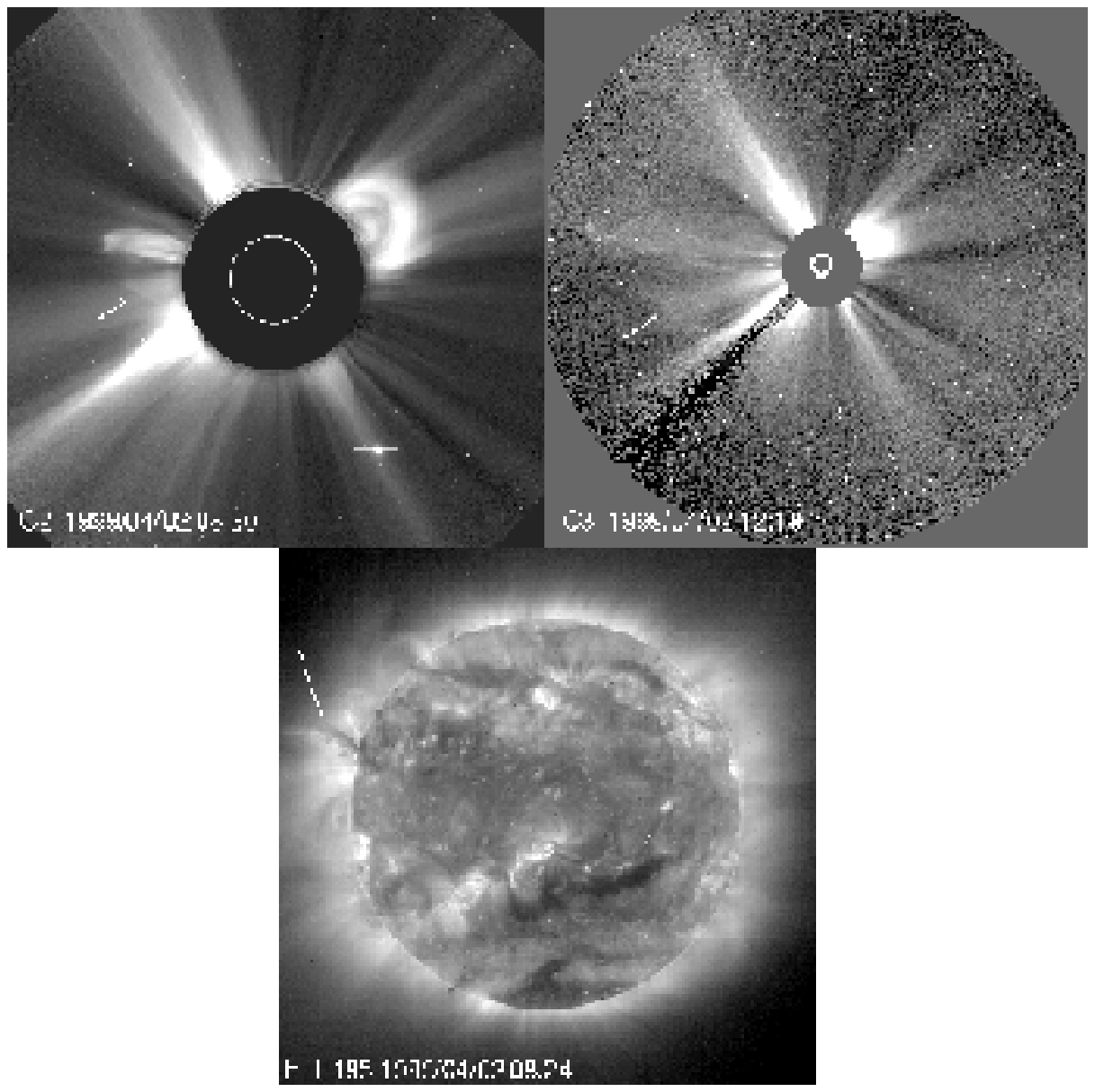}
\figcaption[f1.ps]{LASCO C2 (left), C3 (right) and EIT 195\AA\ (bottom) images
on April 2, 1999, showing the CME event in the overall context of the
solar corona. The arrow in the EIT image shows ejecta that constitute
the core of the white light CME. The arrows, in the C2 and C3 images,
point to the shock-like feature that is analysed in the paper. The
feature is more visible in the processed images in Figure 2. The
circle inside each occulter marks the size of the solar disk for
that coronagraph. Planet Venus is visible in the C2 southwestern
quadrant.}
\end{figure}
\newpage
\begin{figure}
\epsfig{file=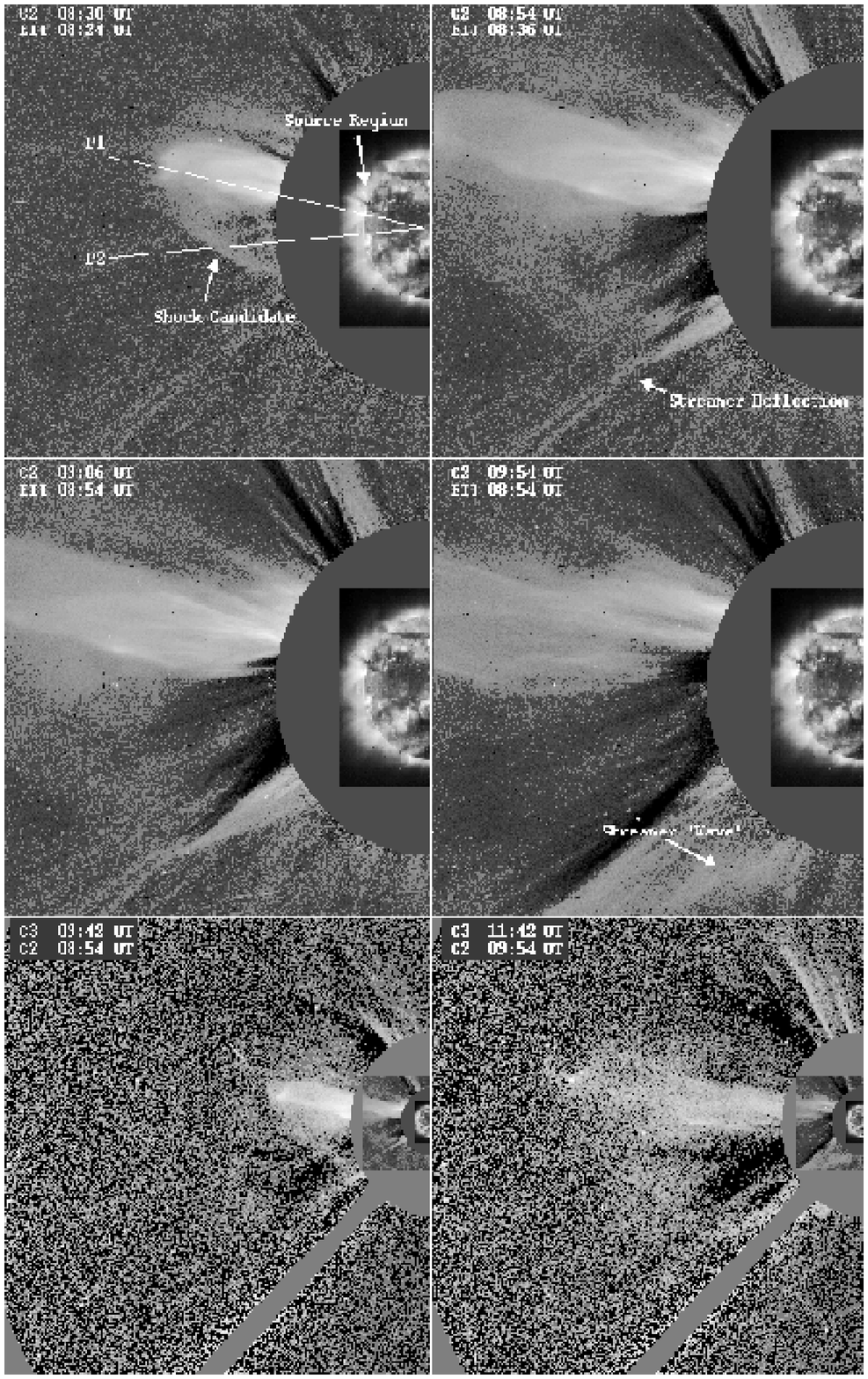}
\figcaption[f2.ps]{LASCO C2, C3 and EIT 195\AA\ observations of the
CME on April 2, 1999. Only part of the eastern field of view is
shown. A pre-event image has been subtracted from the C2 and C3 images
which were subsequently calibrated in units of excess mass (see \S~3.1
for details). The type and time of each observation is shown at the
upper left corner of each frame. The lines labeled P1 and P2 demarcate
the position angles where the height-time and density measurements
were made (see text). Also marked on the images are the locations of
the CME source region, the shock front candidate and other features
discussed in the text.}
\end{figure}
\newpage
\begin{figure}
\epsfig{file=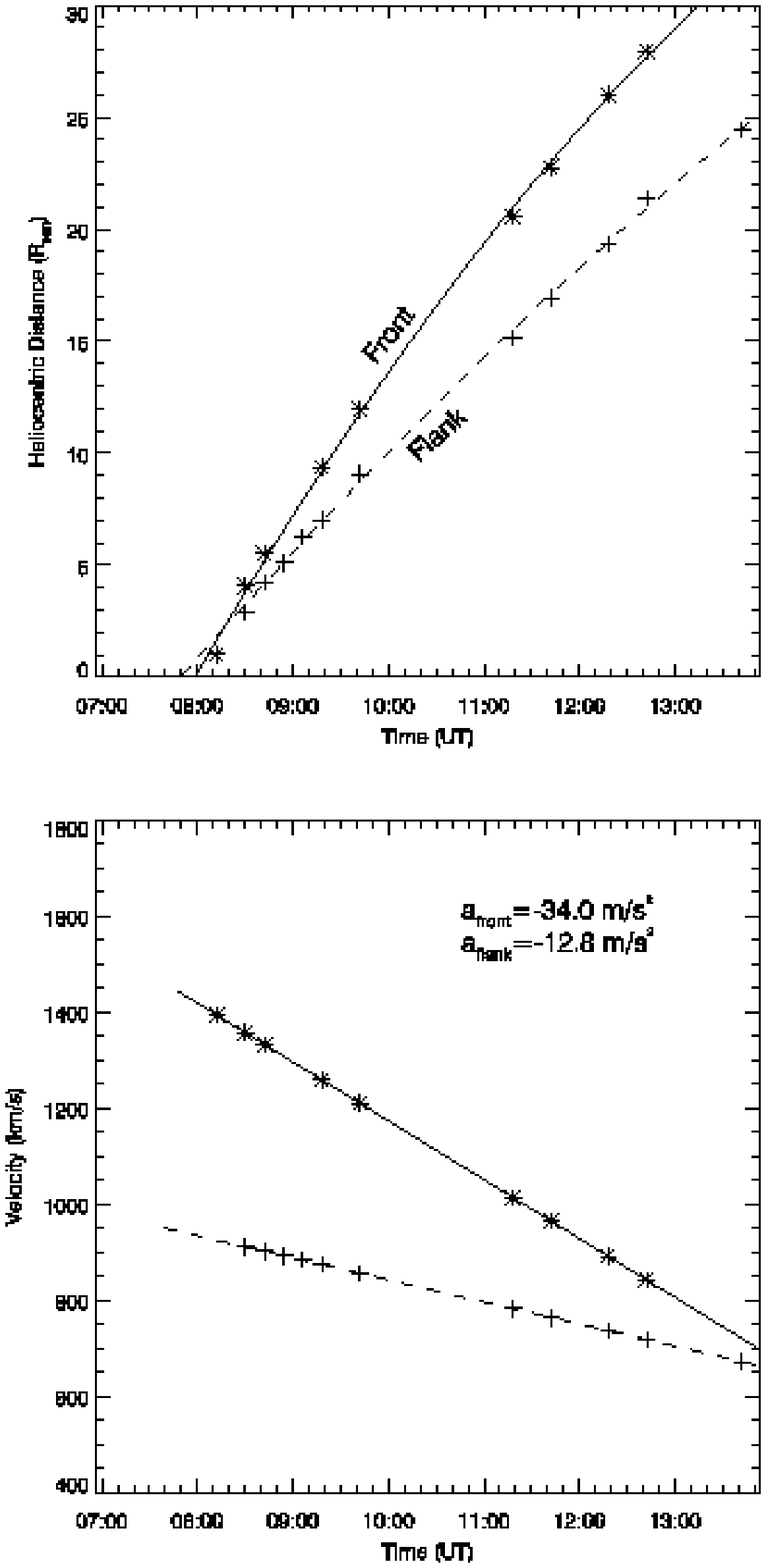}
\figcaption[f3.ps]{Upper panel: Height-time measurements at the CME 
front (stars) and flank (crosses). The lines are second-degree polynomial
fits to the data. Lower panel: The derived speeds and decelerations.}
\end{figure}
\newpage
\begin{figure}
\epsfig{file=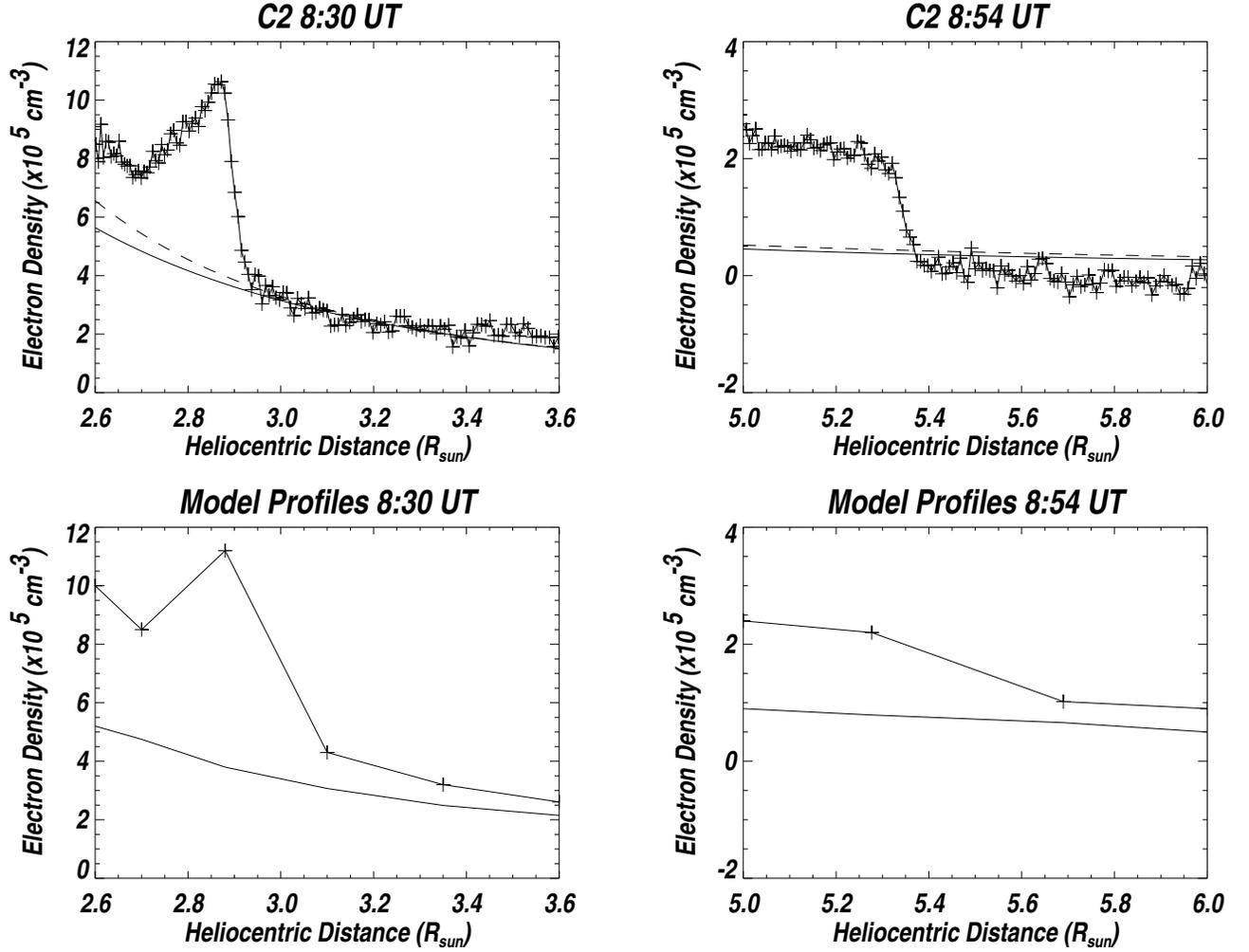}
\figcaption[f4.ps]{(a) Density profiles across the CME flank (lines
with crosses) for two C2 images. The profiles are averages along a
10$^\circ$\ swath perpendicular to the flank normal at P2. The solid
line is the background coronal density derived from a pre-event pB
image.  The dotted line is the SPM model equatorial coronal density
profile. (b) The corresponding simulated density profiles. (see \S~3.2
for details).}
\end{figure}
\newpage
\begin{figure}
\epsfig{file=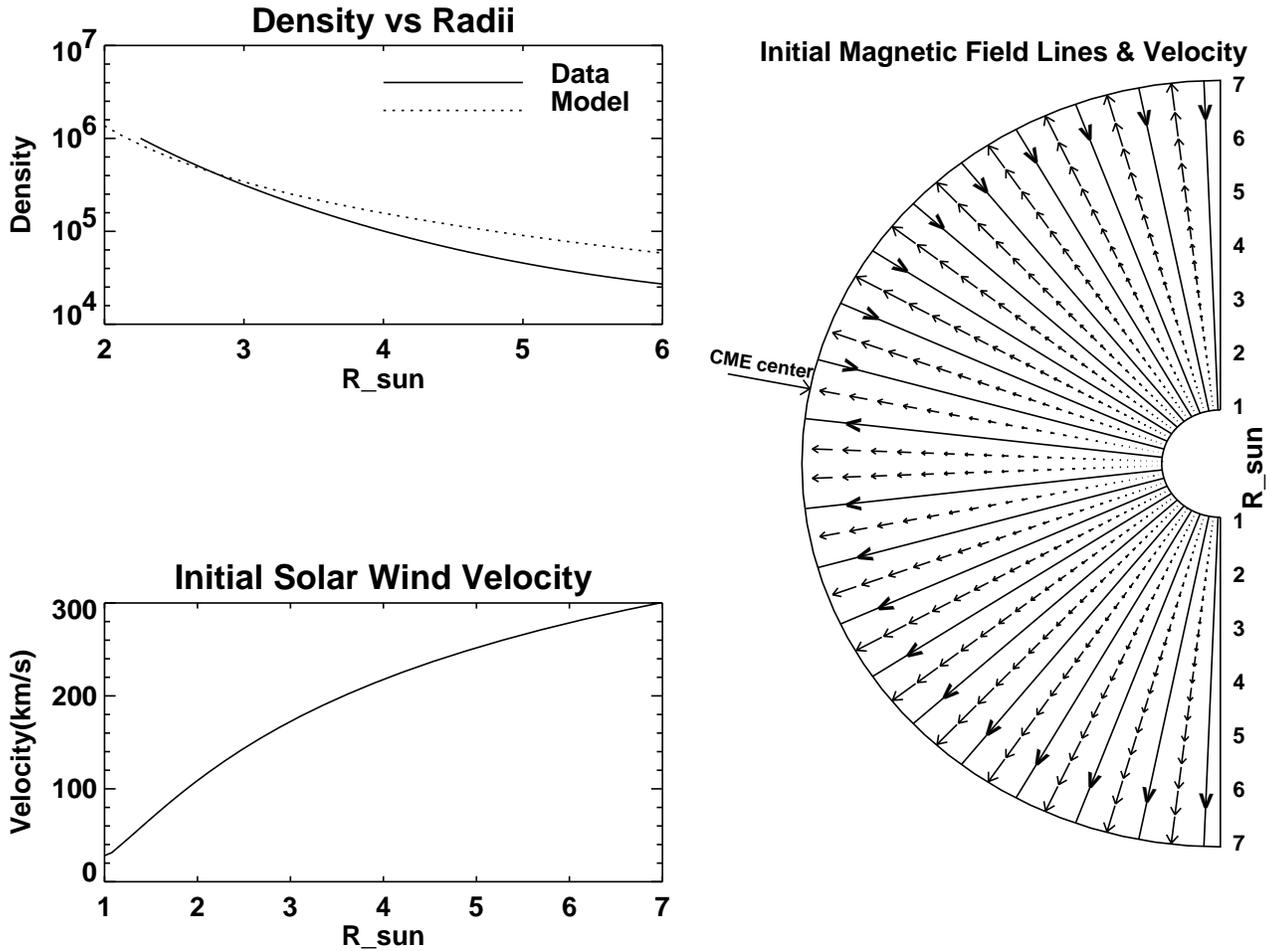}
\figcaption[f5.ps]{The initial quasi-steady state density, background flow
velocity and magnetic field configuration for the simulation. The
radial density distribution of the model is closely matched to the
observed one. In the right panel, the solid lines represent the
magnetic field lines and the dashed lines represent the solar
wind velocity field. The arrows point to the direction of the magnetic 
field and solar wind flow, respectively.}
\end{figure}
\newpage
\begin{figure}
\epsfig{file=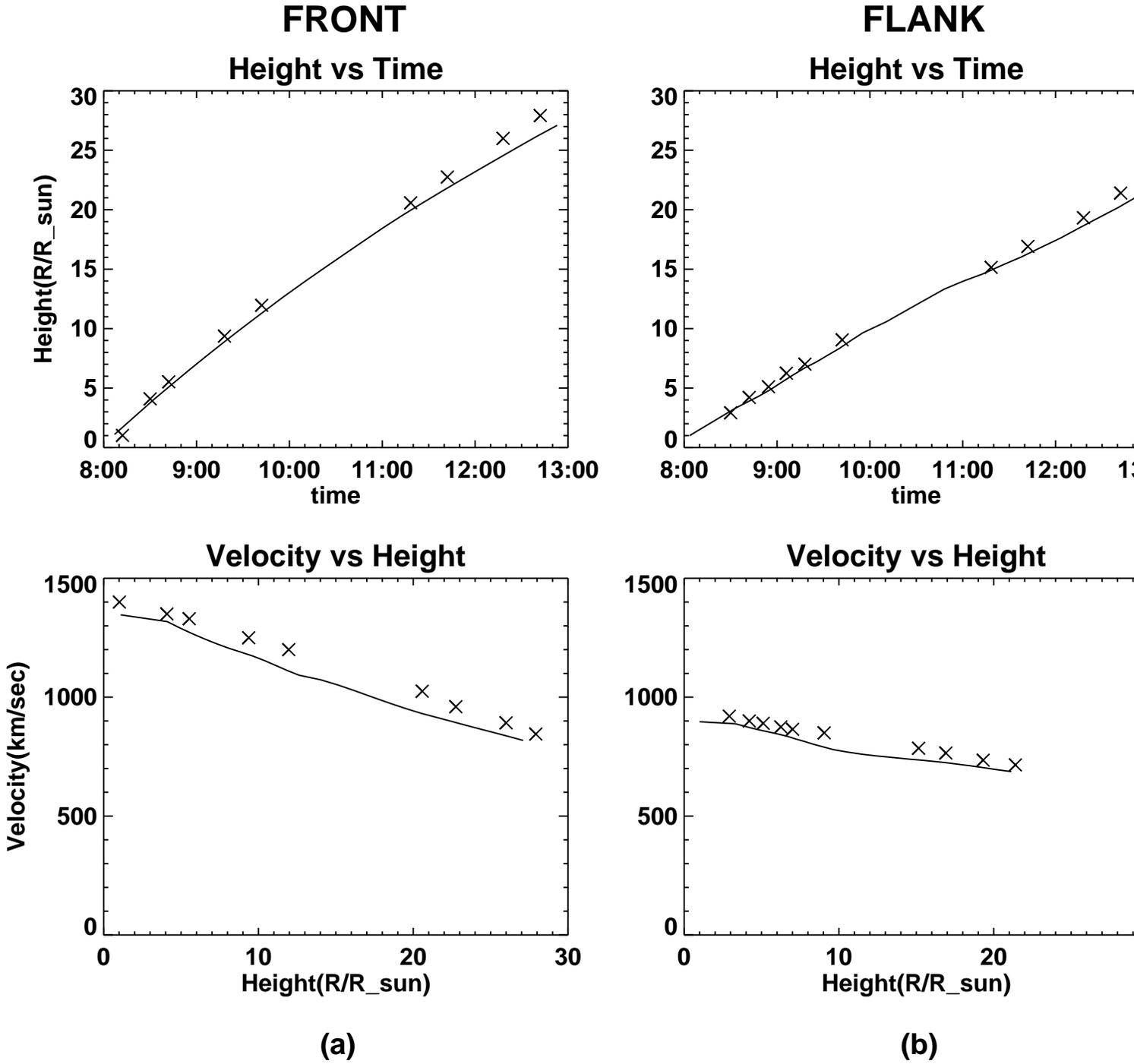}
\figcaption[f6.ps]{Height-time and velocity profiles derived from the MHD
simulation (solid lines) are compared to the LASCO data (crosses) for
both the CME front and flank. The plot is similar to that of Figure
4.}
\end{figure}
\newpage
\begin{figure}
\epsfig{file=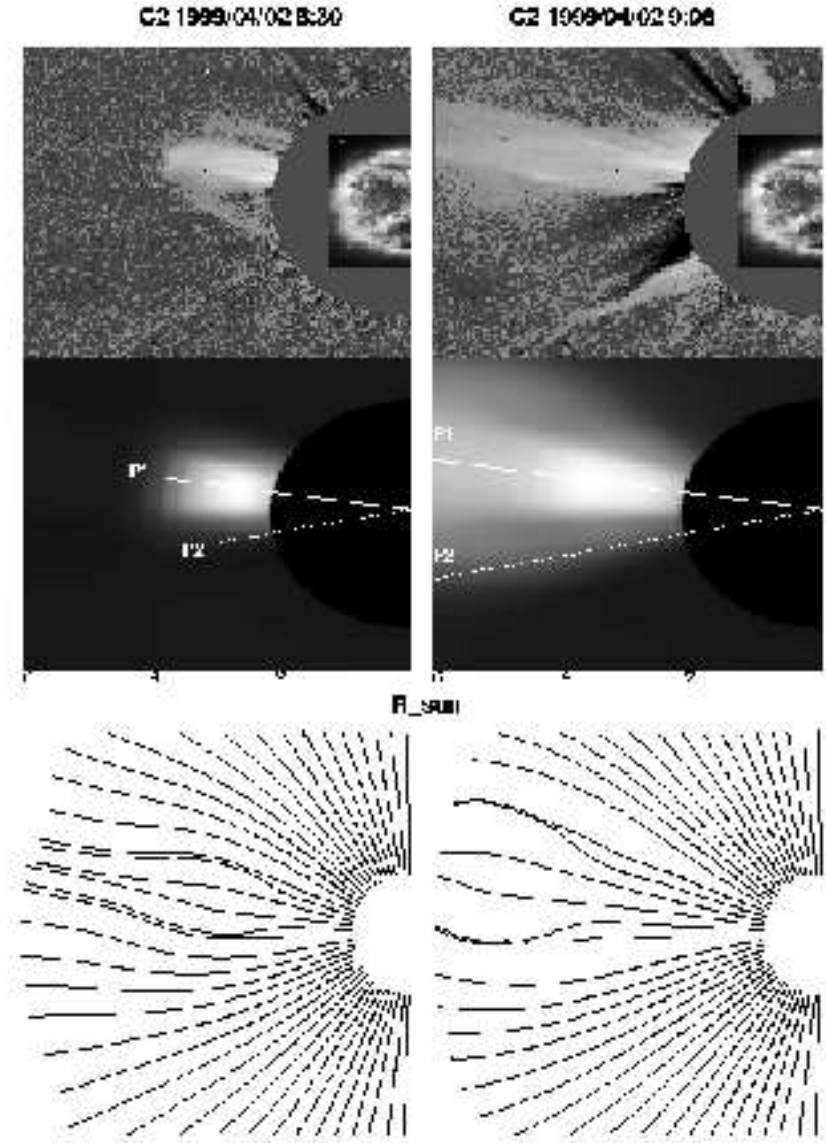}
\figcaption[f7.ps]{Comparison between the C2 observations (top panels)
and the simulated pB images (middle panels) computed from our CME
model. The model magnetic field line configuration is also shown
(bottom panels).}
\end{figure}
\newpage
\begin{figure}
\epsfig{file=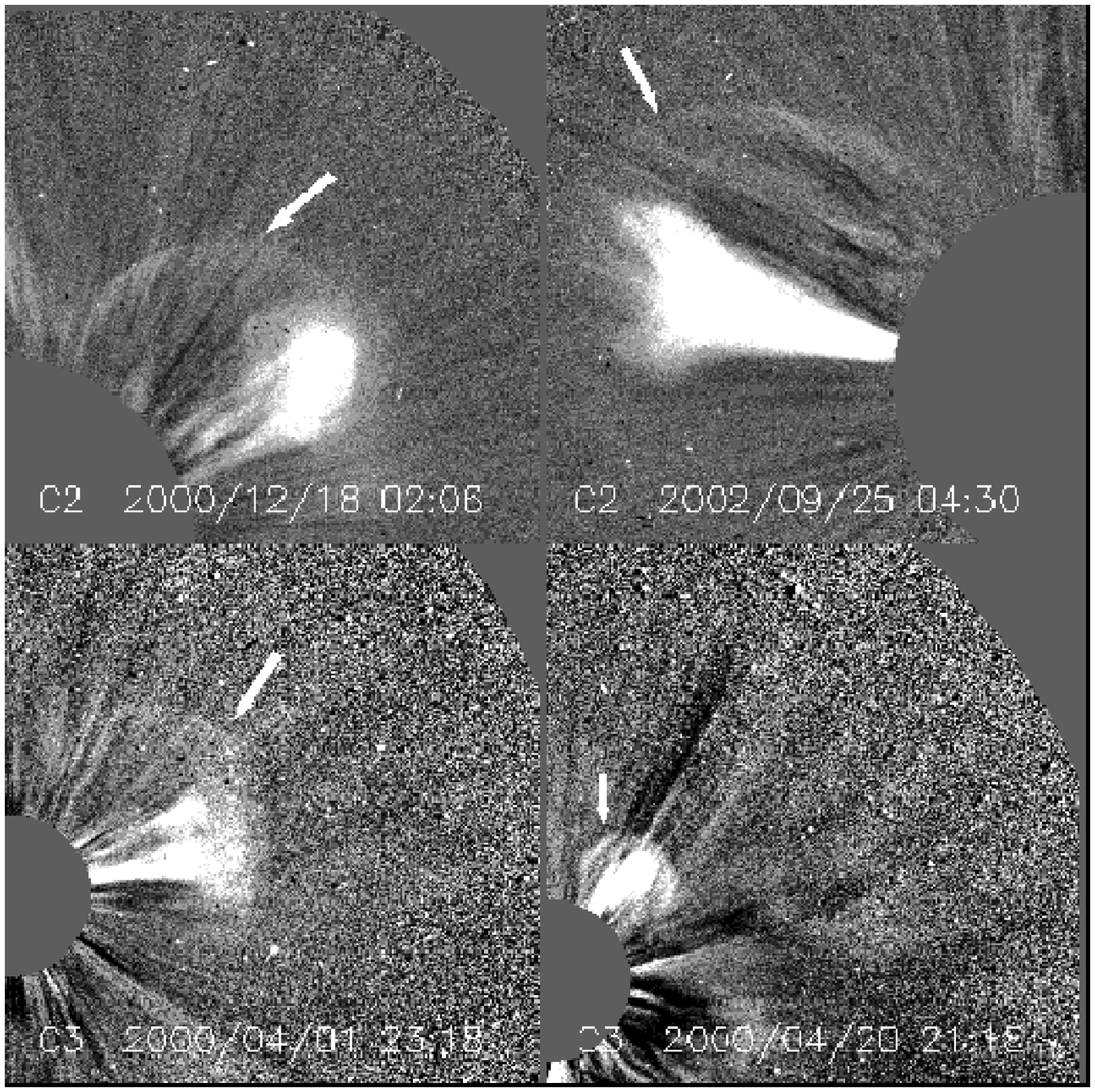}
\figcaption[f8.ps]{Examples of candidate white light shock signatures
in LASCO CME observations. The LASCO telescope and the time of the
event are shown on the figure. The arrows indicate the white light
structure that could be the density enhancement from a shock. A
preevent image has been subtracted from all images.}
\end{figure}

\clearpage
\begin{deluxetable}{ccccccccc}
\tablewidth{35pc}
\tablecaption{The Model-derived Velocity and Shock Data for the
Flank of the CME.} 
\tablehead{
\colhead{Time} & \colhead{$V_s$} & \colhead{$V_f$} &
\colhead{$U_{sh}$} & \colhead{$V_{sw}$} & \colhead{$V_{prop}$} &  
\colhead{$\gamma$} & \colhead{$M_\alpha$} & \colhead{$\rho_2/\rho_1$} \\
\colhead{(UT)} & 
\colhead{(km/s)} & \colhead{(km/s)} & \colhead{(km/s)} & 
\colhead{(km/s)} & \colhead{(km/s)} & \colhead{} & 
\colhead{} & \colhead{} }
\startdata
08:30 & 153 & 348 & 659 & 206 & 865 & 1.05 & 1.89 & 3.36\\
08:54 & 150 & 256 & 588 & 252 & 840 & 1.07 & 2.29 & 4.60\\
09:18 & 147 & 219 & 508 & 317 & 825 & 1.13 & 2.31 & 3.99\\
09:42 & 141 & 206 & 465 & 335 & 800 & 1.22 & 2.25 & 3.61\\
11:18 &  76 & 151 & 302 & 425 & 727 & 1.35 & 2.00 & 2.76\\ 
11:42 &  69 & 134 & 267 & 433 & 700 & 1.37 & 1.99 & 2.71
\enddata
\end{deluxetable}
\clearpage
\begin{deluxetable}{ccccccccc}
\tablewidth{35pc}
\tablecaption{The Model-derived Velocity and Shock Data for the
Front of the CME.} 
\tablehead{
\colhead{Time} & \colhead{$V_s$} & \colhead{$V_f$} &
\colhead{$U_{sh}$} & \colhead{$V_{sw}$} & \colhead{$V_{prop}$} &  
\colhead{$\gamma$} & \colhead{$M_\alpha$} & \colhead{$\rho_2/\rho_1$} \\
\colhead{(UT)} & 
\colhead{(km/s)} & \colhead{(km/s)} & \colhead{(km/s)} & 
\colhead{(km/s)} & \colhead{(km/s)} & \colhead{} & 
\colhead{} & \colhead{} }
\startdata
08:30 & 158 & 532 &1030 & 220 & 1250 & 1.05 & 1.94 & 3.67\\
08:54 & 153 & 334 & 901 & 297 & 1198 & 1.16 & 2.70 & 4.97\\
09:18 & 145 & 302 & 801 & 349 & 1180 & 1.22 & 2.75 & 4.62\\
09:42 & 136 & 284 & 749 & 366 & 1115 & 1.27 & 2.64 & 4.20\\
11:18 &  62 & 172 & 450 & 430 &  880 & 1.42 & 2.61 & 3.39\\ 
11:42 &  46 & 154 & 402 & 438 &  840 & 1.44 & 2.58 & 3.31
\enddata
\end{deluxetable}

\begin{thebibliography}{}
\bibitem[Brueckner et al.(1995)]{brueckner95} 
Brueckner, G. E., et al. 1995, \solphys, 162, 357
\bibitem[Cliver, Webb \& Howard(1999)]{cliver99} 
Cliver, E. W, Webb, D. F., \& Howard, R. A. 1999, \solphys, 187, 89
\bibitem[Delaboudini\'{e}re et al.(1995)]{boudine95} 
Delaboudieni\'{e}re, J.-P. et al. 1995, \solphys, 162, 291
\bibitem[Dobrzycka et al.(2003)]{dobrzycka03} Dobrzycka, D., et
al. 2003, \apj, 588, 586 
\bibitem[Domingo, Fleck \& Poland(1995)]{domingo95} 
Domingo, V., Fleck, B., \& Poland, A. I. 1995, \solphys., 162, 1
\bibitem[Gilbert et al.(2001)]{gilbert01} 
Gilbert, H. R., et al. 2001, \apj, 550, 1093
\bibitem[Gopalswamy et al.(2000)]{gopal00}
Gopalswamy, N. et al., 2000, \grl, 27, 1427
\bibitem[Gosling et al.(1974)]{gosling74} 
Gosling, J. T., et al. 1974, \jgr, 79, 4581
\bibitem[Handy et al.(1999)]{handy99} 
Handy, B. N., et al. 1999, \solphys, 187, 229
\bibitem[Hartle and Barnes(1970)]{hartle70}
Hartle, R. E., \& Barnes, A. 1970, \jgr, 75, 6915
\bibitem[Hayes, Vourlidas \& Howard(2001)]{hayes01}
Hayes, A. P., Vourlidas, A., \& Howard, R. A. 2001, \apj, 548, 1081
\bibitem[Hundhausen, Holzer \& Low(1987)]{hund87} 
Hundhausen, A. J., Holzer, T. E., \& Low, B. C. 1987, \jgr, 92, 11173
\bibitem[Hundhausen(1988)]{hund88} Hundhausen, A. J., 1988, in
Proc. of the 6th Solar Wind Conf., NCAR/TN-306+Proc, p. 181
\bibitem[Hundhausen, Burkepile \& St. Cyr(1994)]{hund94}
Hundhausen, A. J., Burkepile, J. T., \& St. Cyr, O. C. 1994, \jgr, 99, 6543
\bibitem[Jeffrey and Taniuti(1964)]{jeffrey64} 
Jeffrey, A., \& Taniuti, T. 1964, Non-linear wave propagation with 
application to physics and magnetohydrodynamics, Chapter 4, (Academic 
Press Inc: New York)
\bibitem[Klassen et al.(2000)]{klassen00}
Klassen, A., et al. 2000, \aa, 141, 357
\bibitem[Kohl et al.(1995)]{kohl95} 
Kohl, J. L., et al. 1995, \solphys, 162, 316
\bibitem[Maia et al.(2000)]{maia00} 
Maia, D., et al. 2000, \apj, 528, L49
\bibitem[Mann et al.(1999)]{mann99}
Mann, G., et al., 1999, Proc. of SOHO-8 Workshop, Wilson, A. (ed),ESA
SP-446, p. 477
\bibitem[Maxwell and Dryer(1981)]{maxwell81} 
Maxwell, A., \& Dryer, M. 1981, \solphys, 73, 313
\bibitem[MacQueen et al.(1980)]{mcqueen80}
MacQueen, R. M., et al. 1980, \solphys, 65, 91
\bibitem[MacQueen and Fisher(1983)]{mcqueen83}
MacQueen, R. M., \& Fisher, R. 1983, \solphys, 89
\bibitem[Michels et al.(1984)]{michels84}
Michels, D. J., et al. 1984, {\it Adv. Space Res., 4(7)}, 311
\bibitem[Odstr\u cil and Karlick\'y(2000)]{odstr00}
Odstr\u cil, D., \& Karlick\'y, M. 2000, \aa, 359, 766
\bibitem[Poland et al.(1981)]{poland81}
Poland, A. I., et al. 1981, \solphys, 69, 169
\bibitem[Priest(1982)]{priest82}
Priest, E. R. 1982, {\it Solar Magnetohydrodynamics,} 73 pp., (D. Reidel
Publ. Comp: London)
\bibitem[Saito, Poland \& Munro(1977)]{saito77} 
Saito, K., Poland, A. I., \& Munro, R. H. 1977, \solphys, 55, 121
\bibitem[Sheeley, Hakala \& Wang(2000)]{sheeley00} 
Sheeley, N. R., Hakala, W. N., \& Wang, Y.-M. 2000, \jgr, 105, 5081
\bibitem[Sime, MacQueen \& Hundhausen(1984)]{sime84} 
Sime, D. G., MacQueen, M. R., \& Hundausen, A. J. 1984, \jgr 89, 2113 
\bibitem[Sime and Hundhausen(1987)]{sime87}
Sime, D. G., \& Hundausen, A. J. 1987, \jgr, 92, 1049
\bibitem[Spitzer(1962)]{spitzer62}
Spitzer, L., 1962, Physics of Fully Ionized Gases, 2nd rev. ed., (John
Wiley:New York) 
\bibitem[St. Cyr et al.(1999)]{stcyr99}
St. Cyr, O.C., et al. 1999, \jgr, 104, 12493
\bibitem[St. Cyr et al.(2000)]{stcyr00} 
St. Cyr, O.C, et al. 2000, \jgr, 105, 18169 
\bibitem[Steinolfson et al.(1978)]{steinolfson78}
Steinolfson, R. S., et al. 1978, \apj, 225, 259
\bibitem[Steinolfson(1985)]{steinolfson85}
Steinolfson, R. S. 1985, in {\it Collisionless Shocks in the
  Heliosphere: Reviews of Current Research\/}, Geophys. Monogr. Ser.,
vol. 35, B.T. Tsurutani and R. G. Stone (eds), (AGU:Washington, DC), p. 1
\bibitem[Steinolfson(1988)]{steinolfson88}
Steinolfson, R. S., 1988, \jgr, 93, 14261
\bibitem[Steinolfson and Hundhausen(1990a)]{steinolfson90a} 
Steinolfson, R. S., \& Hundhausen, A. J. 1990a, \jgr, 95, 15251
\bibitem[Steinolfson and Hundhausen(1990b)]{steinolfson90b} 
Steinolfson, R. S., \& Hundhausen, A. J. 1990b, \jgr, 95, 20693
\bibitem[Subramanian et al.(1999)]{subramanian99}
Subramanian, P., et al. 1999, \jgr, 104, 22321
\bibitem[Suess, Wang, \& Wu(1996)]{suess96}
Suess, S. T., Wang, A. H., \& Wu, S. T. 1996,\jgr, 101, 19957
\bibitem[Tsuneta et al.(1991)]{tsuneta91} 
Tsuneta, S., et al. 1991, \solphys, 136, 1
\bibitem[Van der Holst, Van Driel-Gesztelyi \& Poedts
  (2002)]{holst02}
Van der Holst, B., Van Driel-Gesztelyi, L, \& Poedts, S. 2002, in
Proc. of the 10th Europ. Sol. Phys. Meeting., Wilson, A (ed), ESA SP-506, p.71
\bibitem[van de Hulst(1950)]{hulst50} 
van de Hulst, H. C. 1950, {\it Bull. Astron. Inst. Netherlands\/}, 11, 135
\bibitem[Vourlidas et al.(2000)]{vour00} 
Vourlidas, A., et al. 2000, \apj, 534, 456
\bibitem[Wang(1992)]{wang92}
Wang, A. H. 1992, Ph.D. dissertation, Univ. of Alabama in
Huntsville. 
\bibitem[Wang et al.(1998)]{wang98} 
Wang, A. H., et al. 1998,\jgr, 103, 1913
\bibitem[Wu et al.(1996)]{wu96} 
Wu, C. C., et al. 1996, \solphys, 165, 377
\bibitem[Wu and Wang(1987)]{wu87} 
Wu, S. T., \& Wang, J. F. 1987,{\it Comput. Meth. Appl. Mech. Eng.\/}, 64,
267 
J. Feynman (eds), (AGU:Washington, DC), p. 83
\bibitem[Wu et al.(1999)]{wu99} 
Wu, S. T., et al. 1999, \jgr, 104, 14789
\bibitem[Wu et al.(2000)]{wu00} 
Wu, S. T., et al. 2000, \apj, 545, 1101

\end{thebibliography}
\end{document}